 \documentclass{emulateapj}

\begin{document}

\title{Spectropolarimetry of C-class flare footpoints}


\author{L. Kleint\altaffilmark{1} }
\affil{High Altitude Observatory / NCAR, P. O. Box 3000, Boulder CO 80307, USA; kleintl@ucar.edu }




\begin{abstract}
      We investigate the decay phase of a C-class flare in full-Stokes imaging spectropolarimetry with quasi-simultaneous measurements in the photosphere (6302.5~\AA\ line) and in the chromosphere (8542~\AA\ line) with the IBIS instrument. We analyze data from two fields-of-view, each spanning about 40\arcsec $\times$ 80\arcsec\ and targeting the two footpoints of the flare.
   A region of interest is identified from $V/I$ images: a patch of opposite polarity in the smaller sunspot's penumbra. We find unusual flows in this patch at photospheric levels: a Doppler shift of -4~km/s, but also a possible radial inflow into the sunspot of 4~km/s. Such patches seem to be common during flares, but only high-resolution observations allowed us to see the inflow, which may be related to future flares observed in this region.
   Chromospheric images show variable overlying emission and flows and unusual Stokes profiles.  
    We also investigate the irregular penumbra, whose formation may be blocked by the opposite polarity patch and flux emergence. The 40 min temporal evolution depicts the larger of the flare ribbons becoming fainter and changing its shape. Measurable photospheric magnetic fields remain constant and we do not detect flare energy transport down from the chromosphere. We find no clear indications of impact polarization in the 8542~\AA\ line. We cannot exclude the possibility of impact polarization, because weaker signals may be buried in the prominent Zeeman signatures or it may have been present earlier during the flare.
\end{abstract}

\keywords{Sun: flares -- Polarization -- Magnetic fields}

\section{Introduction}

Flares are among the most energetic magnetic solar phenomena. Extensive observations are documented in many reviews \citep{aschwanden2002, fletcher2005, benz2008, schrijver2009, hudson2011}, most of them focusing on coronal observations through UV and X-ray measurements, because these are easily accessible with the constant monitoring of spacecraft. But a complete picture of what is happening in all solar layers is still missing, especially simultaneous measurements of magnetic fields and velocities in different layers. It would be desirable to record the full polarization vector quasi-simultaneously in several heights of the solar atmosphere with a high spatial resolution to determine the strength and orientation of the magnetic field vector. It is however still a challenge to observe flares with high-resolution spectropolarimetric imaging instruments such as IBIS or CRISP \citep{cavallini2006, scharmeretal2008} because of seeing conditions, weather limitations, a relatively small field-of-view (FOV) and their unpredictability.

Polarization measurements can be inverted to derive the solar magnetic field vector. The HI\-NO\-DE satellite provides a very useful spectropolarimeter, however only for the two photospheric lines around 6302 \AA. Such measurements have been used to determine the magnetic field configuration and its change during flares \citep[e.g.,][]{murrayetal2011}. But it is an open question what may happen in the chromosphere during these photospheric changes. Other spectropolarimetric flare observations mainly focused on the linear polarization. A linear polarization signal close to the footpoints of a flare may be interpreted as impact polarization generated by an anisotropic distribution of flare particles. But so far, the observations, which were mainly carried out in the H$\alpha$ line,  are contradictory. While some authors report linear polarization signals of up to several percent \citep{henouxetal1990, metcalfetal1992, vogthenoux1999, henouxkarlicky2003} observed with spatial resolutions in the arcsec range, others \citep{biandaetal2005} did not find linear polarization in 30 flares with a sensitivity level of better than 0.1 \%, though with lower spatial resolution (several arcsec). 

Most flares occur in active regions. Emergence of new flux or shear may provide the energy to start a flare. Often, the emergence of flux or a twisted flux rope can be observed in the photosphere hours or days before a flare \citep[][and references therein]{kurokawa1987,brooksetal2003, limetal2010}, creating a patch of opposite polarity close to one of the pre-existing active regions. Shear of the magnetic field may be introduced by a movement of the two spots with respect to each other, if they have a magnetic connection. It is yet unclear if this shearing can cause material to be moved to the corona where the magnetic field may reconnect during a flare or if the flare is triggered by other mechanisms. Photospheric measurements indicate that abrupt changes of the magnetic field strength and structure may occur during flares \citep[][and references therein]{wangetal1994,kosovichevzharkova2001,suetal2011} but it seems that these changes are less pronounced or absent for weaker flares \citep{chenetal1994,petriesudol2010}.

Chromospheric flare observations are much more common than photospheric observations because flares are easily visible in H$\alpha$ and other chromospheric lines. A red asymmetry of chromospheric spectral lines is usually observed after the onset of the flare \citep{svestkaetal1962,ichimotokurokawa1984} and is interpreted as chromospheric material moving downward, which was compressed by heating from above, called the chromospheric condensation. Lines forming higher, for example EUV lines, simultaneously show a blue asymmetry, meaning that material is flowing upwards and increases the density of coronal loops. This chromospheric evaporation probably results in the appearance of coronal loops \citep[][and references therein]{sheeleyetal2004}. 

Our observations try to overcome several of the limitations mentioned above. Our spatial resolution is only slightly lower than HINODE's, enabling us to determine the magnetic field configuration in the photosphere. We have almost simultaneous measurements with a similar resolution in the chromosphere, also recording the full polarization vector and we can search for impact polarization. We can resolve line profiles better than previous imaging measurements, enabling us to look for asymmetries and thus flows at sub-arcsecond resolution. Our time coverage spans most of the decaying phase of a C-class flare, allowing us to study its temporal evolution. This paper focuses on presenting the data, provides a look on interesting features and offers phenomenological explanations for the observations. In a following paper we quantify the results by carrying out NLTE inversions and calculations.


\section{Observations and data reduction}
\label{obs}

The observations were carried out with the IBIS instrument \citep{cavallini2006, reardoncavallini2008} in its spectropolarimetric mode on January~29, 2007. The dual-beam Fabry Perot instrument records "monochromatic" images (see below) and the user may select the wavelengths from several prefilter passbands. 

We recorded 41 wavelength steps in the  \ion{Ca}{2} 8542~\AA\ line and subsequently 26 steps in the \ion{Fe}{1} 6302.5~\AA\ line. During each wavelength step, 6 images of different polarization states were taken ($I$+$Q$, $I$+$V$, $I$-$Q$, $I$-$V$, $I$-$U$, $I$+$U$), which are combined during the data reduction into an image set of the Stokes vectors ($I,Q,U,V$). One cycle of these 402 images took about two minutes. A broad band camera recording the same FOV was run simultaneously to obtain images that reflect the changes in seeing and can be used for speckle reconstructions. 20 scans of the main sunspot of AR 10940 were recorded and afterwards, the telescope was pointed to the smaller spot for another 10 scans. Data from all 30 scans were used where we analyze temporal evolutions and for single images, the scans with the best seeing were selected (Figures \ref{iheight} and \ref{iheightss}, for example).

Chromospheric, but not photospheric brightenings were seen in both sunspots. After a comparison with TRACE images, we determined that we captured both footpoints during the declining phase of a C-class flare. The active region was located at about 4$^\circ$ S and 35$^\circ$ E, resulting in a heliocentric angle $\cos\theta = 0.82$. 

The data reduction of IBIS is not trivial. Apart from the standard steps like dark and linearity correction, flatfielding and alignment of different channels, the reduction also required several other steps. Because of the collimated mount of the Fabry Perot, there is a wavelength variation across the FOV. To obtain true monochromatic images, all spectra have to be interpolated onto a common wavelength scale. Another difficulty is caused by the seeing. The final polarization images are combinations of four single images (dual beam and beam exchange), which are taken at two different times. Speckle reconstructed white light images, calculated with the KISIP code \citep{woegervdl2008}, serve as reference for an automatic program that registers the shifts during each exposure caused by the variable seeing. All images are then interpolated to remove these variations. During several of our scans, the seeing was excellent and a resolution close to 0.33\arcsec\ was reached (plate scale 0.165\arcsec/px). Calibrations for the polarimetric properties of the instrument and the telescope were also performed. A polarimetric sensitivity $< 10^{-2}$ was reached for the fractional polarization states.

A problem with flare data and their large intensity gradients are spurious polarization signals if the data are not perfectly aligned. These signals would be largest close to the flare. Using USAF target images and dot grid images we aligned our data with an automatic correlation routine to an accuracy of better than 0.1~px. For testing, we also introduced deliberate shifts of up to 1~px and the effect was easily visible, reassuring us that our alignment was the best match.

Observations from the TRACE satellite \citep{handyetal1999} were used for complementary information in the wavelengths 195~\AA, 1600~\AA\ and whitelight. The data reduction was performed with the Solarsoft package \citep{freelandhandy1998}. TRACE has a spatial resolution of 1\arcsec\ (0.5\arcsec/px). For the 1600~\AA\ and whitelight filters, TRACE obtained one set of images per 50-102 minutes (variable time intervals) during that day, fortunately one set coincided exactly with the IBIS observing start time. Images with the 195~\AA\ filter were taken in much more frequent intervals of $\sim$1 minute with different exposure times and several blackout periods during the day with variable lengths (around 30 min). 
Hinode observations are not available for this date.

\subsection{Our C3.4 flare}
\begin{figure}
   \centering
    \includegraphics[width=0.5\textwidth]{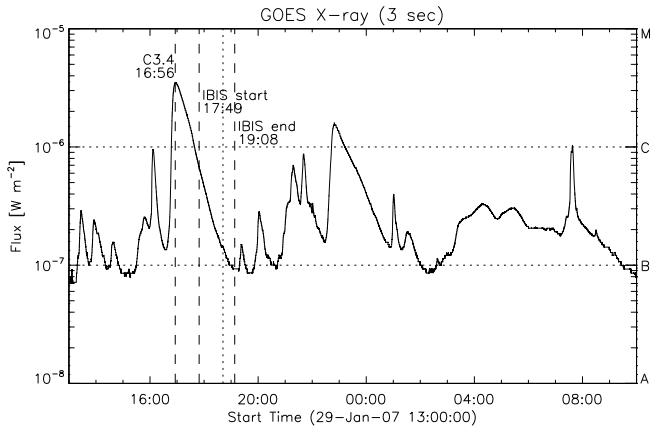}
    \caption{GOES X-ray flux with the IBIS observing times indicated. The vertical dotted line shows when the FOV was changed.
              }
         \label{goes}
   \end{figure}

The flare we observed is classified by GOES as C3.4 event, a rather small flare. It was the strongest event of AR 10940 that produced several C- and B-class flares. The flare started at 16:39~UT, with its maximum at 16:56~UT. IBIS was calibrating during these times and started observing at 17:49~UT. The IBIS observations stopped at 19:08 when the GOES activity was at the B-level but increased shortly afterwards for a series of small flares during the next 12 hours. See Fig.~\ref{goes} for a graphical overview of the X-ray flux.

\begin{figure}
   \centering
    \includegraphics[width=0.45\textwidth]{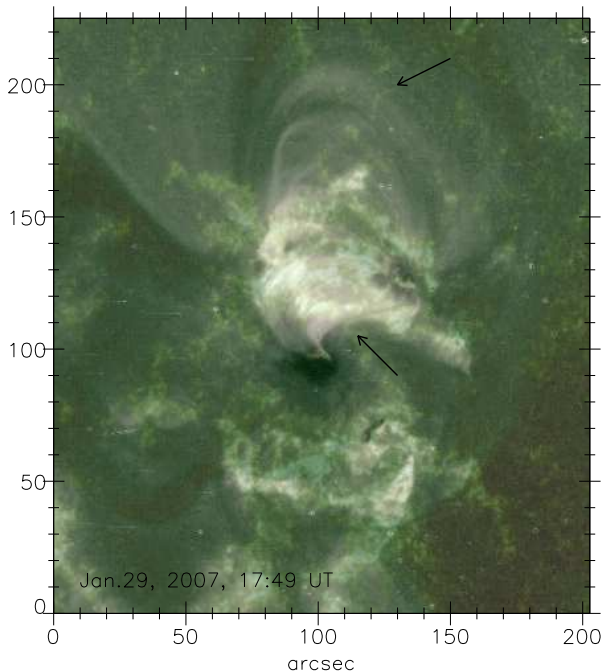}
    \caption{TRACE composite of white light (black, grey),  1600~\AA\ (reddish colors) and 195~\AA\ (green colors and bright loops). The two sunspots in AR 10940 were connected by a loop system (lower arrow) whose footpoints correspond to the bright ribbons observed at chromospheric heights. The upper arrow points to a second loop system connecting the smaller sunspot to several pores. The origin of the coordinate system is arbitrary.
              }
         \label{composite}
   \end{figure}

A composite TRACE image of the active region at 17:49~UT is shown in Fig.~\ref{composite}. Loops are connecting the two ribbons inside the two sunspots, which are shown in high-resolution in the next section.


\section{Results}

\subsection{Intensity with height}

\begin{figure*}
   \centering
     \includegraphics[width=\textwidth]{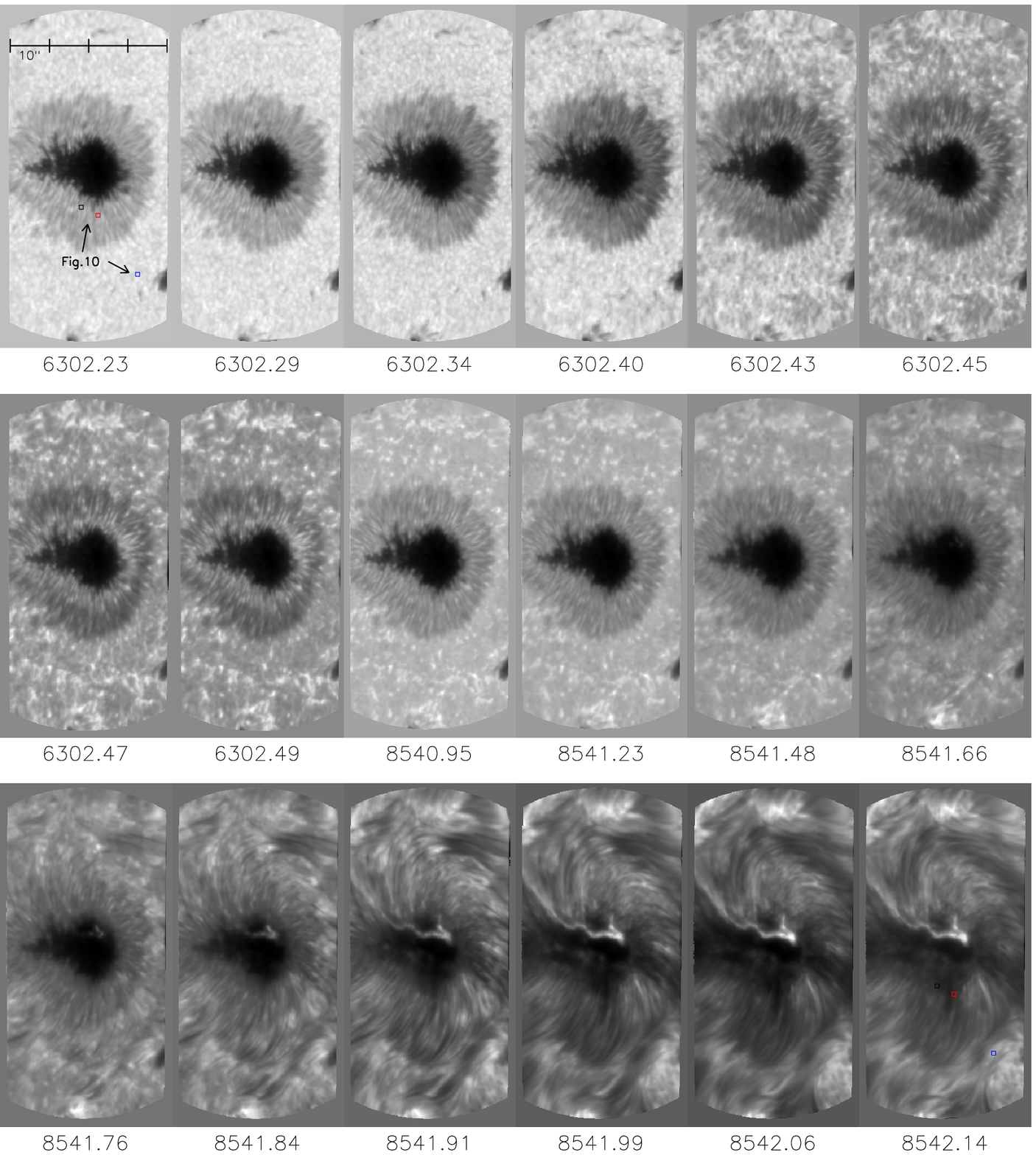}
     \caption{Images of selected wavelength steps during a scan of two spectral lines targeting the main spot of AR 10940 at 18:10-18:12~UT. The images range from the lower photosphere (top left) to the lower chromosphere (bottom right). The wavelengths below each image are given in \AA. The FOV of each image is about 40\arcsec $\times$ 80\arcsec. A flare ribbon starts appearing above a certain formation height (8541.66~\AA). The colored boxes show where the spectra for Fig.~\ref{set3} were obtained.       }
         \label{iheight}
   \end{figure*}

\begin{figure*}
   \centering
     \includegraphics[width=\textwidth]{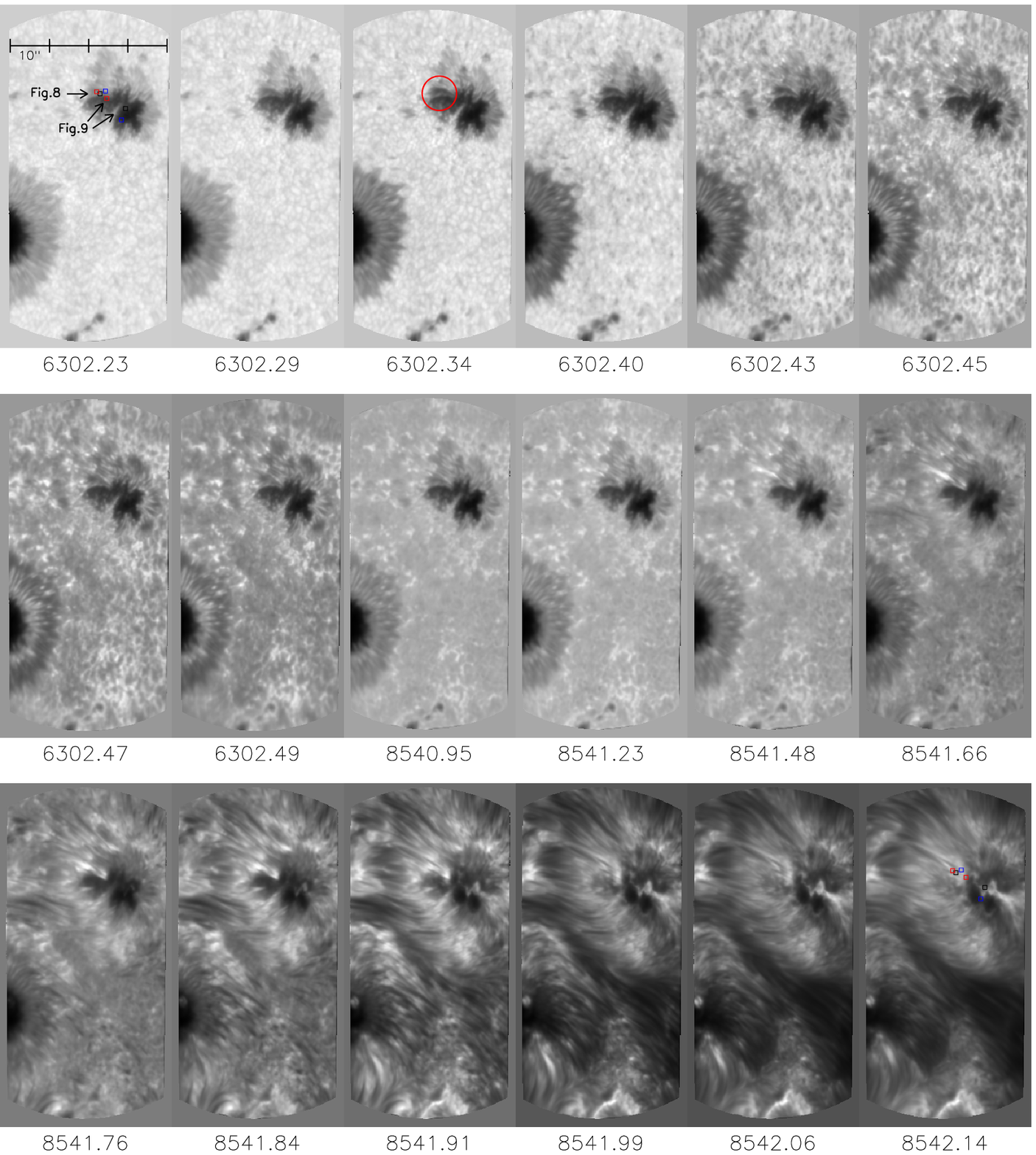}
     \caption{Images of selected wavelength steps during a scan of two spectral lines targeting the smaller spot of AR 10940 at 18:42-18:44~UT. The wavelengths below each image are given in \AA. Note the irregularly shaped small spot, a patch in its penumbra with slightly different intensity (circled), which corresponds to strong emission in wavelengths around 8541.84~\AA, and the appearance of a flare ribbon inside the small sunspot at chromospheric heights.}
         \label{iheightss}
   \end{figure*}

Figures \ref{iheight} and \ref{iheightss} show a selection of wavelengths from two different scans. The FOV of the second figure was shifted by about two heliographic degrees to the west and one degree to the north. The wavelength and thus the height in the solar atmosphere is increasing towards the right and the bottom of the figures. The image axes coincide with the usual solar reference frame (E to the left) and disk center is about 35$^\circ$ towards the west.

It can be seen that the line core of \ion{Fe}{1} (6302.49~\AA) shows similar features as the line wing of \ion{Ca}{2} (8540.95~\AA), although the contrast is better in the line core image. Probably this is because of scattering in the source function of Ca due to NLTE.

At the formation height range that corresponds to about 8541.7~\AA, a bright ribbon starts appearing in both sunspots. These ribbons are the footpoints of a coronal loop system connecting the two sunspots. One of the ribbons is located inside/above the umbra of the bigger spot and the other one seems to cross the smaller spot extending into its penumbra. TRACE images show another loop system (upper arrow in Fig.~\ref{composite}) connecting the smaller spot ribbon to an area of pores or proto-spots in the east, outside of the IBIS FOV.

\subsection{Opposite polarity patch}
\label{polarity}

\begin{figure*}
   \centering
     \includegraphics[width=\textwidth]{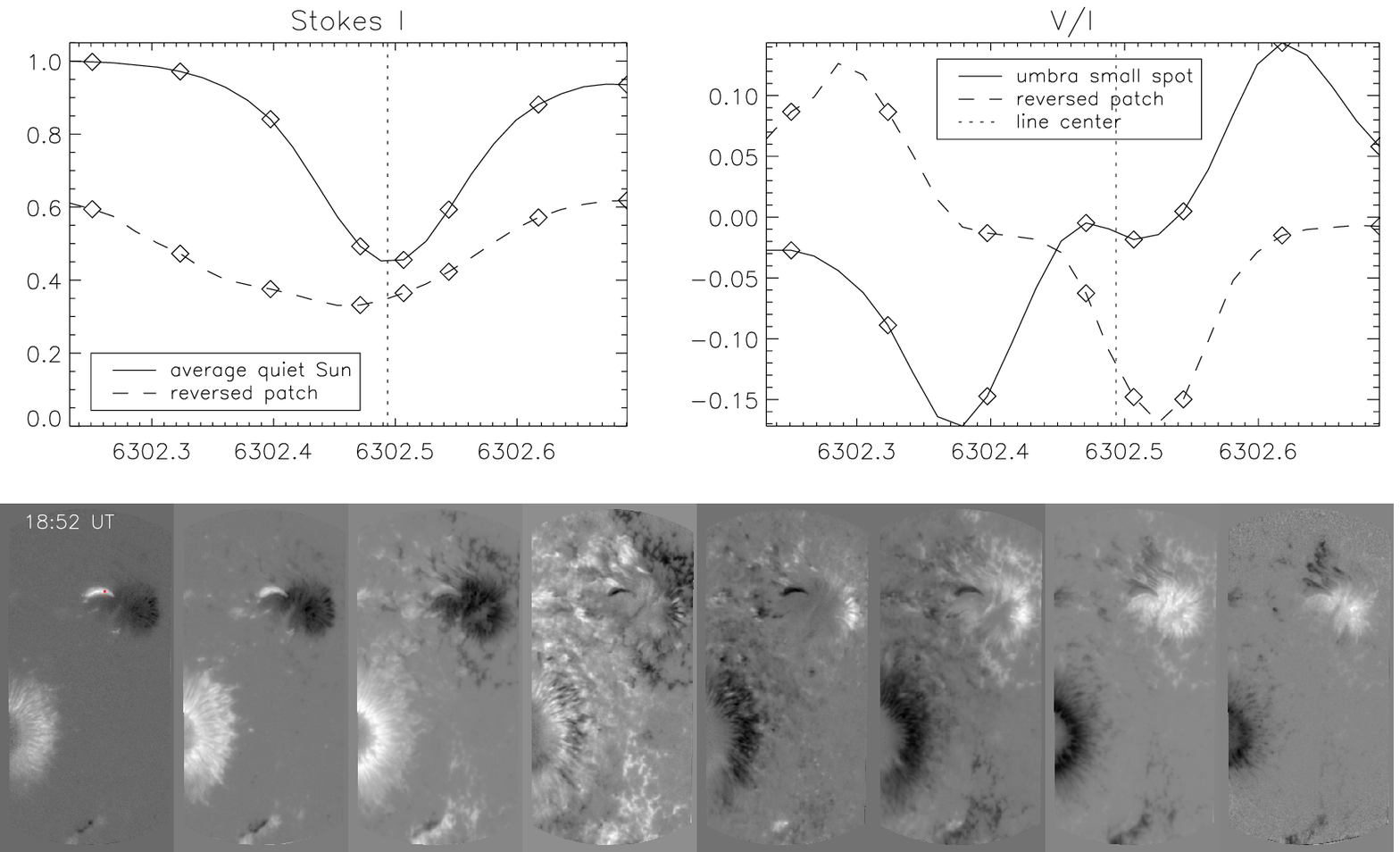}
     \caption{Analysis of a small opposite polarity patch in the penumbra of the small sunspot. \textit{Top left:} Intensity profiles for the average quiet Sun (solid) and an area of 0.5\arcsec $\times$ 0.5\arcsec\ inside the OPP, marked by a red dot in the lower left image. \textit{Top right:} Stokes V/I profiles for the umbra of the small spot (solid) and the OPP (dashed). The diamonds in all plots denote the wavelengths of which a V/I image is shown at the \textit{bottom}. The vertical dotted lines denote the line center in case of no Doppler shifts.}
         \label{polrevers}
   \end{figure*}

A small patch of the penumbra of the smaller spot (circled in Fig.~\ref{iheightss}) is noteworthy. Its length is about 5.6\arcsec and it has opposite polarity to that sunspot and a strong photospheric Doppler shift towards the blue. Fig.~\ref{polrevers} shows the patch as white elongated structure in the first two images in the bottom. The bottom panel represents selected V/I images at different wavelengths of the 6302~\AA\ line. The plot on the top left shows two line profiles, normalized to the continuum level of the quiet Sun: a regular, quiet Sun profile (solid) and a profile averaging an area of about 0.5$\arcsec \times$ 0.5\arcsec\ inside the opposite polarity patch (OPP; dashed). The quiet Sun profile was determined from an average over $8\arcsec \times 25\arcsec$ of granulation west of the bigger spot and the profile did not change significantly for half or double of this box size. The smaller box size for the OPP was chosen in order not to average intensity profiles with different Doppler shifts and its size and location is shown as tiny red dot in the first $V/I$ image of the figure. The patch shows an intensity level of less than 0.6 I$_c$, which is slightly darker than the rest of the penumbra, and the line is broadened significantly. Such a line profile may be expected if there are several unresolved upflowing components. The top right plot shows two V/I profiles: a regular, umbral profile from the small spot (box size 0.5$\arcsec \times$ 0.5\arcsec) with a zero-crossing at the dotted line, which means zero Doppler velocity. The dashed V/I profile is again from the same area inside the OPP. It is clearly fully reversed and shifted towards the blue wavelengths. The maximum blueshift we find inside the patch corresponds to an up- or inflow of about -4 km/s. The whole patch shows Doppler shifts towards the blue, also at its edges.

Speckle reconstructed white light images can be used for correlation tracking of flows. The whole OPP shows a possible radial inflow, while other penumbral filaments show the regular Evershed flow (radial outflow). Darker dots can be seen to move from the granulation at the eastern tip of the patch to the western tip, vanishing inside the sunspot. Correlation tracking of different features reveals velocities of about 4 km/s inwards. This correlation tracking was done manually by identifying several small dark features in consecutive speckle images (total 44 speckle images with time intervals of 27 s) and calculating their velocities from their measured pixel positions.
It is unclear if this is a true inflow or possibly moving footpoints of magnetic field lines or tiny downflows. 

The chromospheric images of the OPP are much harder to interpret. As can be seen in Fig.~\ref{iheightss}, there is emission at that location. Thus, one has to be careful to determine whether a $V/I$ profile is reversed because of emission or a different polarity. The emission is variable, often appearing and disappearing within less than two minutes. Sudden brightenings can also be observed mainly in the blue line wing, indicating large Doppler velocities. There certainly are places where photospheric and chromospheric polarities coincide (=i.e. opposite polarity to sunspot), but generally, the chromosphere shows mixed polarities in this location and highly irregular Stokes profiles (see Fig.~\ref{set1} for examples). 

\subsection{Penumbra formation}

\begin{figure*}
   \centering
     \includegraphics[width=\textwidth]{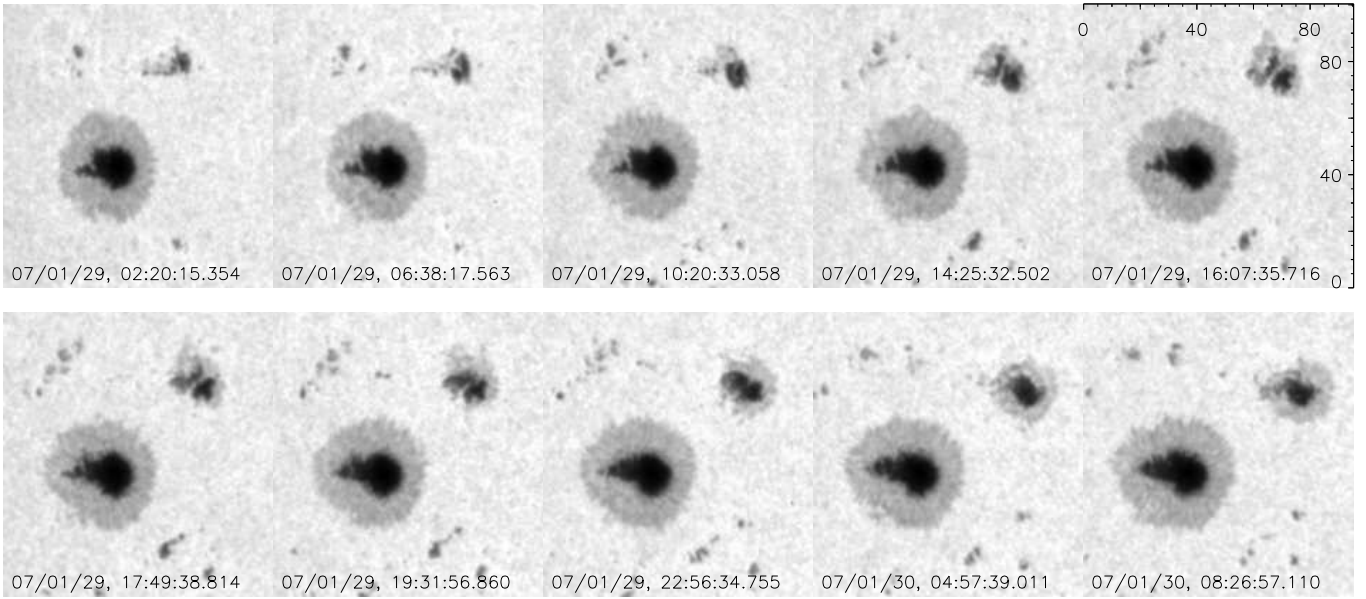}
     \caption{Temporal evolution of the active region NOAA 10940 during 30 hours observed by TRACE. Several B- and C-class flares happened during this time, all of them invisible in these white light images. The smaller sunspot is rotating clockwise and a penumbra is forming. The image scale is given in arcseconds and IBIS observations were taken during the time between the first two images in the bottom row.}
         \label{wlevol}
   \end{figure*}

Figure~\ref{wlevol} depicts the temporal evolution of our observed active region as seen by TRACE in white light. The first image shows the situation 15 hours before we started observing and the smaller sunspot does not have a penumbra yet. The penumbra forms during the day and similar to the observations of \citet{schlichenmaieretal2010},  the penumbra formation seems to be suppressed towards the opposite polarity sunspot. It seems that the OPP in the smaller sunspot is the spatial dividing place between penumbra and no penumbra. Only after 4:57 on January~30, a full, although slightly irregular, penumbra develops, which coincides with declining flare activity. Unfortunately, we did not find any high-resolution observations to determine if the OPP still existed at that point (the next Hinode magnetogram was taken on January 31 at 10:55 and did not show the OPP anymore). The smaller sunspot seems to rotate clockwise during the TRACE observations, while the bigger sunspot's orientation remains constant.

\begin{figure}
     \includegraphics[width=0.5\textwidth]{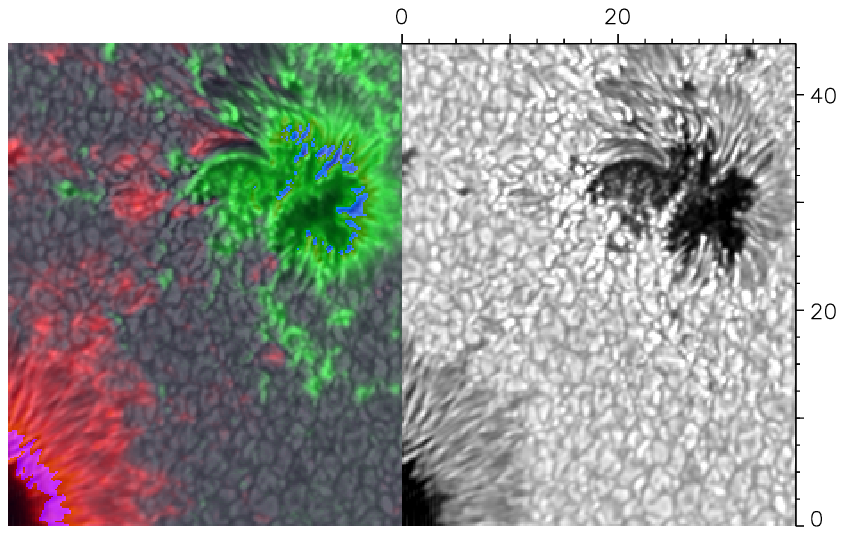}
     \includegraphics[width=0.34\textwidth]{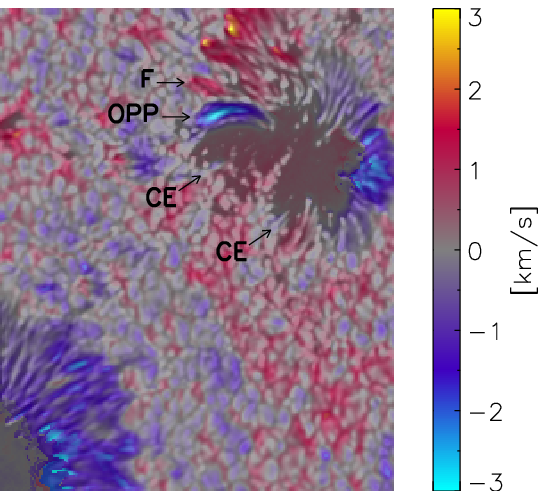}
     \caption{\textit{Top left:} Stokes $V$ image at 6302.38~\AA\ overlaid on top of a speckle reconstructed intensity broadband image. The polarization image shows a combination of Doppler shifts and magnetic fields, which cannot be disentangled in this single frame, in green/red. Most prominent are the opposite polarity patch, the spots and some mixed polarities left of the smaller spot.
     \textit{Top right:} The same speckle reconstructed intensity image for comparison. \textit{Bottom:} Dopplergram overlaid on top of the speckle image. Up- and downflows (blue / red) were clipped at $\pm$ 3 km/s. Images taken at 18:52 UT.}
   
         \label{specklecomp}
   \end{figure}

\citet{schlichenmaieretal2010} pointed out elongated granules in their observations, located in the axis of the active region and attributed them to flux emergence, consistent with numerical simulations \citep{cheungetal2008}. We also find elongated granules, predominantly east (left) of the polarity reversal in the small sunspot, where the flux emergence region is located in our case.  The locations of the mixed polarities mostly correlate with the elongated granules but are not limited to them. These features are shown in Fig.~\ref{specklecomp}. The top left image is a composite of a Stokes $V$ image away from the line core (6302.38 \AA; green/red) and a speckle reconstructed intensity image (greyscale), shown again on the right. Because Doppler shifts shift the Stokes $V$ profiles, the strongest signal does not necessarily represent the strongest magnetic field. The composite clearly shows that mixed polarities occur mainly east of the smaller spot and seem to coincide with irregularly shaped granulation. 

Our time series enables us to visually inspect the flows. A photospheric Dopplergram, determined from bisectors of the line core of Stokes $I$ is shown in the bottom of Fig.~\ref{specklecomp}. The bisectors were determined for each $I$-profile at two levels (4 and 14 \% line depth) and the averaged Doppler shift of these two levels with respect to the quiet Sun line core position was used for the Dopplergram. The Evershed flow is a radial outflow along the penumbra, which can be seen from Dopplershifts around sunspots that are not at Sun center. The limb-side of the sunspot generally shows a redshift and the center-side a blueshift. In our observations, the expected Evershed flow occurs around the bigger spot and the western side of the smaller spot. However, the eastern side of the smaller spot is interesting. The filament (F in Dopplergram) right above the OPP shows the regular radial outflow (red in Dopplergram), but as previously mentioned, the whole OPP shows a possible radial inflow of 4~km/s. This inflow might also prevent the formation of the penumbra close to its location by influencing the geometry (inclination) of the magnetic field lines. \citet{schlichenmaieretal2010b} found a rather strong and yet unexplained counter-Evershed flow of more than 1.5~km/s before the penumbra appeared. We find a similar behavior in two areas on the eastern side of the small spot (CE in Dopplergram) in addition to the OPP. Each of these areas is about 2\arcsec\ in length and they both exhibit flows opposite (blue in Dopplergram) to the expected Evershed flow. Their velocities of  $\sim$0.5~km/s are however significantly lower than those of the previously observed counter-Evershed flows and we cannot exclude unresolved components with different relative speeds, especially because the Stokes $V/I$ profiles are irregular at these locations.

\subsection{Unusual Stokes profiles}

Because of the rich variety of line profiles, especially for the 8542~\AA\ line, there is no simple solution to obtain Dopplergrams or magnetograms. Particularly, places with multiple magnetic components or asymmetric line profiles would lead to misleading interpretations. We therefore present some examples of Stokes profiles in this section and provide a qualitative explanation, while further analyses will be carried out with inversions in a forthcoming paper.

\begin{figure}
   \centering
     \includegraphics[width=0.5\textwidth]{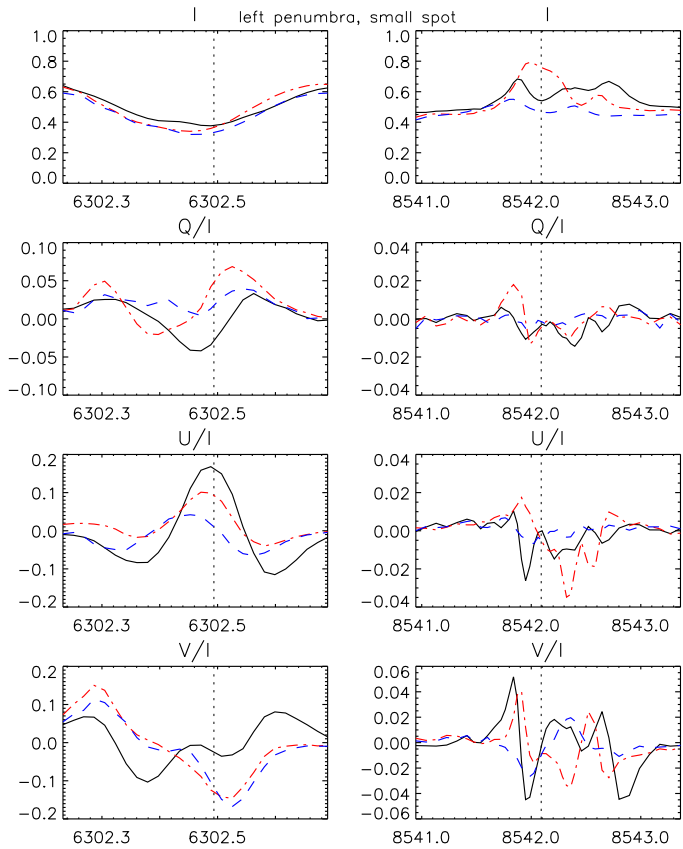}
     \caption{Stokes profiles at different locations, each averaged over an area of 0.33\arcsec $\times$ 0.33\arcsec. \textit{Left:} Photospheric 6302~\AA\ line, $I$, $Q/I$, $U/I$, $V/I$ (top to bottom), showing different areas (different colors and line styles) near the small sunspot inside and close to the opposite polarity patch (see Fig.~\ref{iheightss} for color-coded locations).
     \textit{Right:} Profiles of the same areas in the 8542~\AA\ line. Note the emission in $I$ and the complex structures in the fractional polarization states. 18:51-18:52 UT.}
         \label{set1}
   \end{figure}

\begin{figure}
   \centering
     \includegraphics[width=0.5\textwidth]{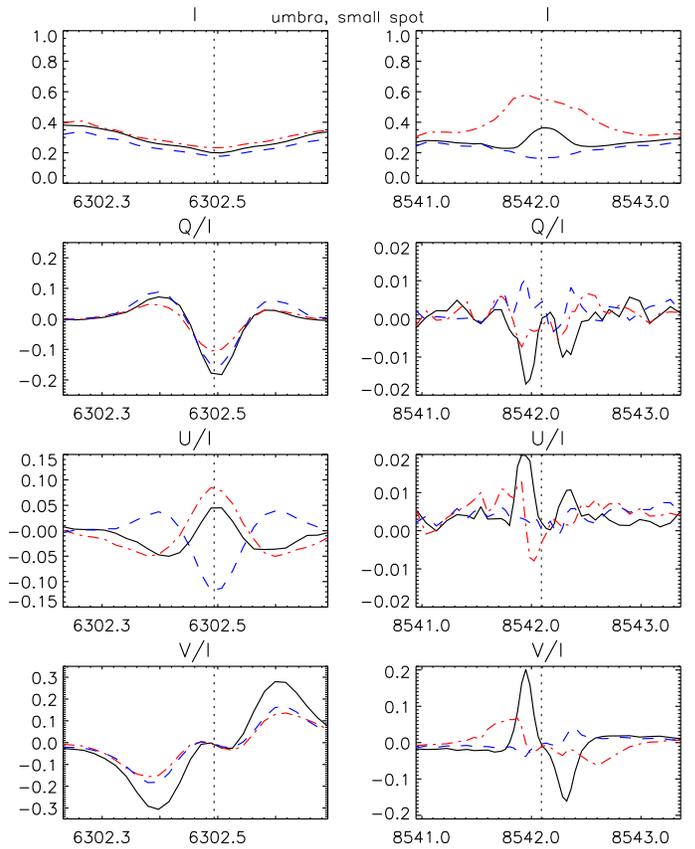}
     \caption{Similar to Fig.~\ref{set1}. These profiles show areas inside the smaller sunspot's umbra. Solid black line: Area inside flare ribbon, just west of photospheric light bridge. Dash-dotted red line: area 5.2\arcsec\ towards the opposite polarity patch, inside flare ribbon. Dashed blue line: 3\arcsec\ from first area towards south, outside flare ribbon (see Fig.~\ref{iheightss} for color-coded locations). 18:51-18:52 UT.}
         \label{set2}
   \end{figure}

\begin{figure}
   \centering
     \includegraphics[width=0.5\textwidth]{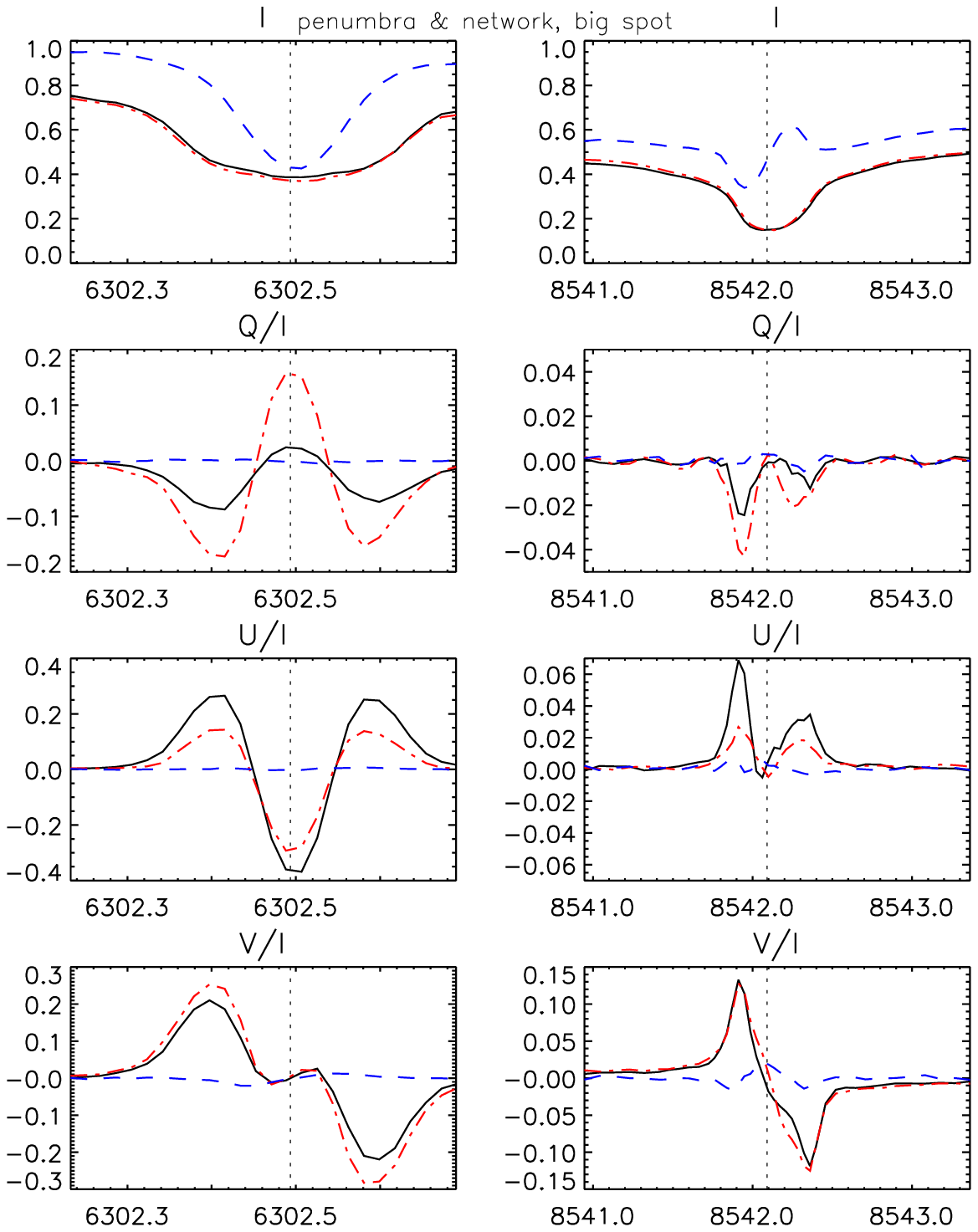}
     \caption{Similar to Fig.~\ref{set1}. Solid and dash-dotted lines show profiles in the southern penumbra of the bigger sunspot. The dashed line shows network south-west of the bigger spot (see Fig.~\ref{iheight} for color-coded locations). Note the little variation of field orientation with height in this part of the penumbra and the unusual 8542 $I$ profile for the network. 18:10-18:12 UT.}
         \label{set3}
   \end{figure}

Figures \ref{set1} - \ref{set3} show the Stokes profiles for different areas (0.33\arcsec $\times$ 0.33\arcsec) of our FOVs. The left column presents the photospheric profiles and the right column the chromospheric profiles ($I$, $Q/I$, $U/I$, $V/I$, from top to bottom). The locations where the profiles were measured are shown as color-coded boxes in Fig.~\ref{iheight} and \ref{iheightss} in a photospheric and a chromospheric image (the box sizes are magnified by a factor of 1.5 for better visibility). The locations were chosen to represent a variety of profile shapes that are common in the given areas. Note that the polarization scale is different for each panel to show maximum details.

Penumbral profiles of the smaller sunspot are shown in Fig.~\ref{set1}. The solid profiles are from a location next to the OPP, the dashed and dash-dotted profiles from inside the OPP. The photospheric profiles show Doppler shifts, multiple components and asymmetries. The chromospheric profiles give an idea why it is so complicated to derive physical quantities. The emission profiles in 8542~\AA\ Stokes $I$ are very broad and seem to consist of multiple components. Stokes $Q/I$ is close to the noise-level, but Stokes $U/I$ shows asymmetric peaks, completely unrelated to the profile shapes in the photosphere. Stokes $V/I$ indicates magnetic elements of the same polarity with large velocities with respect to each other (max.~$\sim 30$~km/s from the zero-crossings). The conclusion from these profiles seems to be that the magnetic fields in the photosphere and the chromosphere at this location are unrelated in strength and orientation. However, because the most unusual Stokes profiles in the chromosphere are seen around these locations, coinciding with the unusual OPP in the photosphere, there seems to be some connection.

Umbral profiles of the smaller sunspot are shown in Fig.~\ref{set2}. The photospheric profiles are as expected from textbooks. The field azimuth is variable within the umbra (similar $Q/I$, but different $U/I$ profiles) and the strong polarization of $V/I$ indicates a strong LOS field. The central reversal of the $V/I$ profiles is probably because of magneto-optical effects \citep{wittmann1971}. The emission in the chromospheric $I$ shows that two profiles (solid, dash-dotted) are from inside the ribbon in the umbra. One of them (solid line) shows a weak transverse Zeeman effect, with the field orientation again seemingly unrelated to the photospheric profiles. Such very weak linear polarization profiles, even in a sunspot's umbra, pose another problem to deriving chromospheric magnetic fields accurately. The chromospheric $V/I$ shows that the same polarity persists with height in the umbra, as expected. Reversals of the profiles are solely because of emission in Stokes $I$.

Fig.~\ref{set3} shows example profiles from the southern part of the bigger spot's penumbra (solid and dash-dotted) and from a network element (dashed blue). The penumbral profiles show similar features in both lines, meaning that the magnetic field orientation seems to be similar, unlike in most of the other shown examples. This does not necessarily imply a connection of field lines throughout the atmosphere, but we think there is a simpler explanation. Obviously the penumbral magnetic field is aligned with the penumbral filaments in the photosphere. When we now focus on the orientation of the chromospheric filaments (cf.~Fig.~\ref{iheight}) around the bigger sunspot, we notice that they are aligned with the penumbra in many places around the bigger spot, but more twisted around the smaller spot. The chromospheric magnetic field is aligned with these filaments in the first order \citep{jaimehector2011}, which explains the similar photospheric and chromospheric field orientations in this part of the FOV. Note that the $\pi$-component of the linear polarization profiles is strongly reduced in 8542. The network profile is shown because of its curious chromospheric $I$ profile. Such small reversals in either the red or the blue part of the line core seem to be common in bright filaments.

\subsection{Impact polarization}
If an anisotropic distribution (e.g., a beam) of particles impacts lower levels of the solar atmosphere  during a flare, this may lead to linear polarization - the impact polarization \citep[for its definition see e.g.,][]{henouxetal1990}. 

Several authors have reported linear polarization up to several percent \citep[e.g.,][]{henouxetal1990, metcalfetal1992, vogthenoux1999, henouxkarlicky2003}, which is above the noise level of our observations (less than 1\% for the fractional polarization states). The reported linear polarization signals peaked after the impulsive phase. However, the study with the highest polarimetric sensitivity so far (one magnitude better than ours and than most of the previous studies) but with relatively low spatial resolution (several arcsec) by \citet{biandaetal2005} did not show any linear polarization in more than 30 flares. The causes for the discrepancy may range from instrumental effects, such as spurious signals when large intensity gradients are present or insufficient spatial resolution to detect the linear polarization, to physical reasons if the impacting energetic particles do not reach the formation height of the observed line or if the particle beam is isotropic. Our spatial resolution is better by about one magnitude than that of previous studies and if there are signals above 1\% they should be visible in our data.

In the smaller sunspot ribbon, there is no linear polarization signal that could be attributed to impact polarization. Only very minor Zeeman signals of up to 2 \% are visible in $Q/I$ and $U/I$.

\begin{figure}
   \centering
    \includegraphics[width=0.5\textwidth]{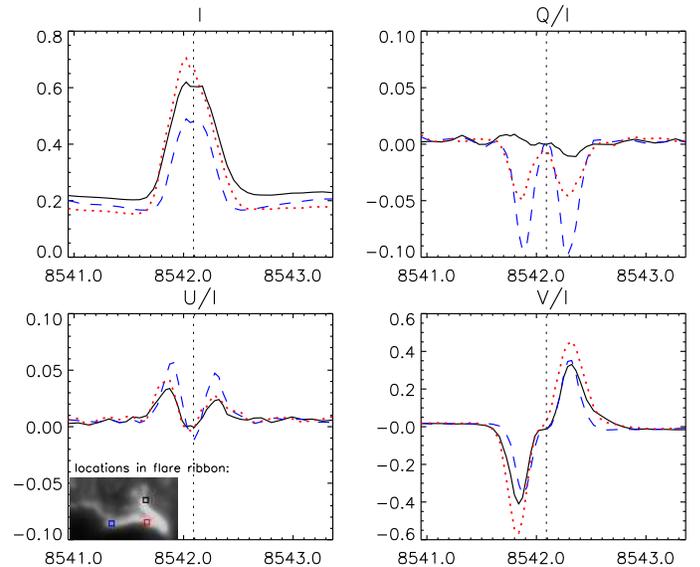}
    \caption{Example of Stokes profiles ($I, Q/I, U/I, V/I$ from top left to bottom right) at three different locations in the flare ribbon, which are depicted in the image below $U/I$. Typical signs of Zeeman polarization signals can be seen, but no impact polarization is visible. Images taken at 18:03 UT.}
         \label{impactpol}
   \end{figure}

The bigger sunspot ribbon does show linear polarization signals. However, they are dominated by the typical symmetric Zeeman shape (second derivative of $I$) and probably not due to impact polarization, which would be proportional to $I$. Figure \ref{impactpol} shows the spectra of the Stokes components in three randomly chosen locations along the ribbon of the bigger sunspot. The spectra are averages over 0.33\arcsec $\times$ 0.33\arcsec, and the image below the $U/I$ plot shows the color-coded locations where the spectra were measured. A linear polarization signal up to 10\% can be seen, but a clear impact polarization signal is missing. The strength of some of these linear polarization signals varies by a few percent during the 40 minutes of data taking, but their shapes remain similar.
  
It is possible that impact polarization was present earlier during the flare or that significant anisotropy might not have reached the formation height range of our spectral line, because of the weakness of the flare. All we can conclude is that during this declining phase of a C-class flare there was no impact polarization visible in 8542~\AA.

\subsection{Flare ribbons}
Flare ribbons are a mainly chromospheric phenomenon corresponding to the footpoints of the coronal loops. It is generally believed that their source of energy are either impacting particles which are accelerated as reconnection occurs in the corona or thermal conduction \citep{asaietal2004}. 

\begin{figure}
   \centering
    \includegraphics[width=0.5\textwidth]{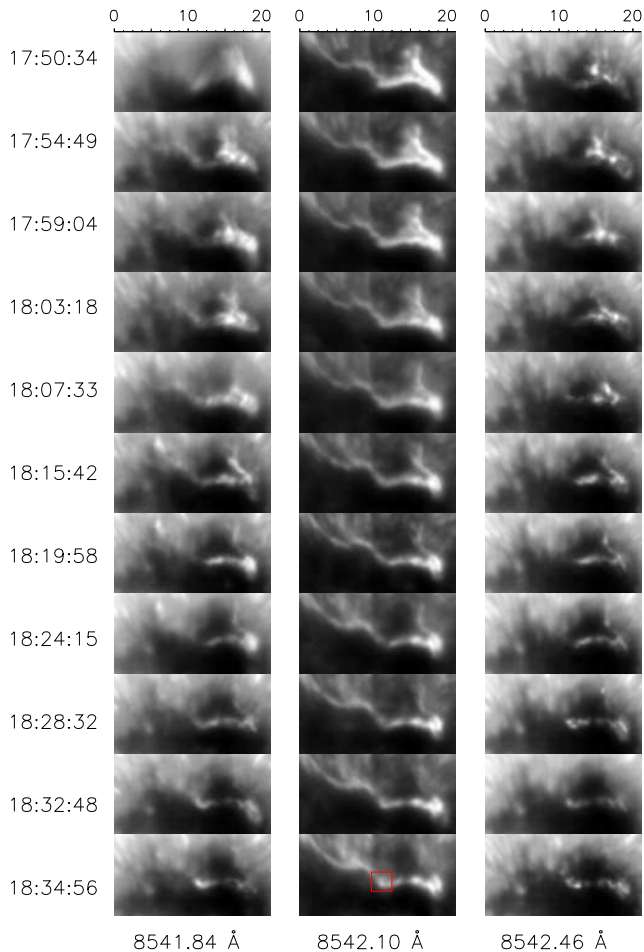}
    \caption{Temporal evolution of the flare ribbon in the umbra of the bigger sunspot. Selected observing times are denoted on the left of three images showing the blue wing, the line core and the red wing of the \ion{Ca}{2} 8542 line. Generally, a decrease of the intensity can be observed with time, apart from small local enhancements (for example inside the red box).
              }
         \label{tempevolribbon}
   \end{figure}

\begin{figure}
   \centering
    \includegraphics[width=0.5\textwidth]{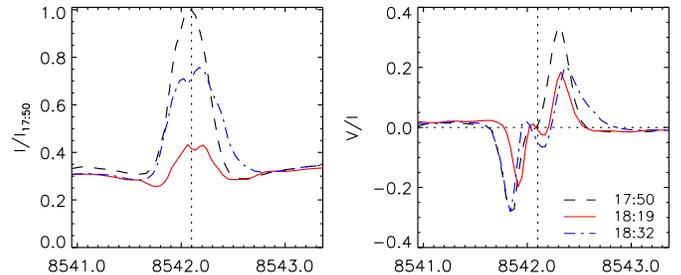}
    \caption{Stokes $I$ (left panel) and $V/I$ (right panel) profiles at three different observing times. They show the evolution of a 0.33\arcsec $\times$ 0.33\arcsec\ patch (inside the red box in Fig.~\ref{tempevolribbon}), whose intensity first decreases and then suddenly increases.
              }
         \label{tempevolpt4}
   \end{figure}

The temporal evolution of the flare ribbon in the bigger sunspot is shown in Fig.~\ref{tempevolribbon}. The three columns show the temporal evolution in the blue wing (left), the line core (middle) and the red wing (right) of the 8542~\AA\ line. The line core images show the weakening of most of the ribbon until only a bright line remains. The upper part of the ribbon vanishes completely, probably because of cooling. Most Stokes $V/I$ signals in the ribbon in the 8542~\AA\ line decrease roughly by a factor of two during our observing time (similar to what can be seen in Fig.~\ref{tempevolpt4}), which certainly is related to the declining intensity, but only inversions may show if the actual field strength is decreasing.
Most of the 8542~\AA\ profile shapes remain constant. The photospheric images of that area, in which flare ribbons cannot be seen, do not indicate any changes in intensity or field strength. 


Interestingly, some small parts of the ribbon become brighter with time, for example the area marked by a red box. The line core intensity of that part decreased monotonically until 18:24 to less than 40\% of its initial value (I$_{17:50}$). But within 8 minutes, its intensity doubled again and reached 0.75$\cdot$ I$_{17:50}$ as is shown in the left panel of Fig.~\ref{tempevolpt4}. This behavior is certainly not an instrumental effect because the photon counts of the line wings remained constant during these times. The $V/I$ amplitudes (Fig.~\ref{tempevolpt4}, right panel) dropped from 0.34 to 0.22 from 17:50 --18:15, while the profile had the typical antisymmetric shape (dashed profile). Two minutes later, the profile started to become asymmetric as if two magnetic components were present, also visible in Stokes $I$ (red solid line). When the intensity started increasing, the line-of-sight magnetic field ($V/I$) was decreasing, though the red lobe of $V/I$ was enhanced. The enhancement continued during our observations (blue dash-dotted $V/I$ profile). The asymmetric profiles may be explained by a downflowing component with a speed of 4.9~km/s and an upflowing component with 6.3~km/s, derived from the Doppler shifts of the zero crossings of the Stokes $V/I$ profiles.

\section{Discussion and conclusions}
\subsection{Opposite polarity and emerging flux}
It has been speculated that emerging twisted flux ropes may trigger or contribute to triggering solar flares \citep[see review of][]{schrijver2009}. These ropes may create a strong-gradient polarity-inversion line (SPIL) which seems to be a prerequisite for M- and X-class flares. Our observations show that even small C-class flares may require such a polarity inversion, though on smaller scales and only visible with high enough resolution. Several possibilities on the actual flare triggering have been suggested by \citet{brooksetal2003}: reconnection of the emerging flux may occur, either with its surrounding magnetic field or the overlying coronal field or a part of the flux rope may be separated and could then rise into the corona increasing its free energy. The emerging flux may increase the shearing, thus destabilizing the magnetic field, which may lead to an eruption. 

If the observed motions are a true radial inflow, this may be a hint of how the next flare might be triggered: magnetic fields of opposite polarity accumulate inside the smaller spot and once a critical limit is reached, reconnection may occur and either send plasma towards the corona or destabilize the whole field geometry enough to cause a flare. This observation is in line with the reconnection scenario suggested by \citet{brooksetal2003}.

This OPP is reminiscent of the observation of \citet{limetal2010}, who used HINODE data to analyze the emergence of a different flux thread with opposite polarity to its surroundings (sunspot and penumbra) before an X-class flare. They also found Doppler shifts in their structure, though with significantly lower velocities (up to -1~km/s) than what we observed. They also noticed downflows up to 2.0~km/s at both tips of the thread which we do not see in our structure. This may be an effect of the different resolutions of the instruments, or different solar dynamics. In their Fig.~3 it can be seen that the patch persisted for at least 3 hours after a X3.4 flare occurred and even seemed to get larger after the flare. The temporal evolution of our patch,  about $\sim$20 minutes of data, does not show any measurable changes in size, strength or structure in all photospheric measurements. The chromospheric region above this patch is highly variable, especially the emission in the blue 8542~\AA\ line wing. Our higher measured velocity may also have a connection to the ongoing flare activity (4 flares during the next 6 hours).

Because of the emission, the high velocities, the complex Stokes profiles and the variable strength of $V/I$ in the chromosphere, it is possible that flux emergence and cancellation occurred in the chromospheric layers of this patch, possibly caused by the underlying unusual photospheric field configuration, again supporting the theory that flares are triggered by processes in lower atmospheric layers. 
\subsection{Flare ribbons}
Because the ribbons are absent in photospheric images, we can assume that they only appear higher than the upper photosphere in this flare. The images of the ribbons in the inner 8542 line wings show even more variability than those from the line core. Brightenings seem to move along the ribbon with time or appear suddenly at certain locations. It is unclear if the hotter ribbon plasma really moves spatially by convection or other forces or if the evolution is driven by events in the corona, possibly without spatially moving the chromospheric plasma. 

The standard flare model states that when the reconnection site in the corona rises, the loops become bigger and thus the chromospheric footpoints separate. Our observed sudden brightenings may be caused by accelerated electrons when a new arcade reconnects. The footpoints are located in a strong magnetic field (as opposed to plage regions, for example) and we do not observe any spatial motion, consistent with most previous observations and models \citep{fletcherhudson2001, lizhang2009}.

This scenario should imply that a correlation between local brightenings in the line core images and a brightening in the line wing should exist. While that is true in a small amount of places, generally the line core intensity is decreasing in spots where the ribbons become visible in the line wing images. It is possible that accelerated electrons reach these lower levels first and then the heat evaporates to higher levels (chromospheric evaporation). Sudden brightenings might also be caused by local reconnection. One of the important questions is where the energy for the observed changes comes from and what may drive the evolution of the ribbons.  

So what may be happening in the flare ribbons? In order to emit light they require constant heating, because the radiative cooling time at this height is about 1.5 minutes \citep{giovanelli1978}. Because the photosphere does not show any changes, we can assume that the heating comes from above (i.e.~the corona). Local brightenings could either be caused by enhanced particle beams from the corona or by local changes such as reconnection inside the ribbon. For particle beams, one may expect redshifted material, similar to the red asymmetry during the impulsive phase \citep{svestkaetal1962}. With our measurements, we cannot determine which scenario is more likely. As the energy release (brightening) occurs, the apparent flux decreases. After a few minutes, the density in the chromosphere may get high enough to introduce motions of the magnetized plasma. This may explain the up- and downflowing components in the observations.

The ribbon in the smaller sunspot shows less variability. Compared to the big spot ribbon (BSR), its average intensity is weaker by about 25\%. In general, its intensity is decreasing, though not as rapidly as in the BSR. Several locations show temporary increases in intensity, but maximally by a factor 1.2, unlike a factor 2 in our example in the BSR. The shape of the small spot ribbon (SSR) remains constant during the 20 minutes of data taking. Stokes $V/I$ also shows only small changes and most of the profiles only decrease by a few percent during the observations, while their shapes remain constant. An exception are times, when $I$ is increasing: In these cases the $V/I$ profiles become asymmetric as if multiple magnetic components were present, however the changes in the $V/I$ amplitudes are very small, consistent with only small changes in Stokes $I$. We think that we may start to see the variable electron precipitation at this resolution.

The small spot ribbon (SSR) appears at slightly lower optical depths in the solar atmosphere than the big spot ribbon (BSR). While first traces of the ribbon become visible at 8541.66~\AA\ in the BSR, they appear at 8541.76~\AA\ for the SSR. This is consistent with the current picture of flares where more energetic particle beams reach deeper levels in the solar atmosphere. It would be worthwhile to observe ribbons during M- and X-class flares and to determine their depths depending on the strength of the flare.

\subsection{Conclusions}
We have analyzed the first high-resolution (0.165\arcsec/px) spectropolarimetric dataset of the footpoints of a C-class flare in the photosphere and in the chromosphere. The following conclusions are obtained:\\
- A small, 5.6\arcsec\ wide patch with opposite polarity is seen in the smaller spot's penumbra in the photosphere. Such patches have been observed in connection with larger flares but it seems that they are also present during smaller flares. Doppler shifts persist inside the patch with velocities of up to -4~km/s. Correlation tracking shows material apparently moving from the granulation into the sunspot with velocities of about 4~km/s inside the patch. This may lead to shearing or reconnection and contribute to triggering flares. Indeed, in the chromosphere, this patch is the region with the largest temporal variations (flux emergence and intensity brightenings). \\
- A penumbra is developing around the smaller sunspot. While it has earlier been observed that penumbra formation is suppressed towards the opposite polarity spot, we find that the opposite polarity patch seems to be the natural division between a part with penumbra and a part without. The development of a full penumbra coincides with declining flare activity.\\
- Highly irregular and asymmetric Stokes profiles render simple calculations of Doppler- and magnetograms impossible. Multiple components with relative flows seems to be common, especially around the flare footpoints. The chromospheric 8542~\AA\ line shows small reversals in its intensity profiles in parts of the 'quiet Sun' and strong emission profiles in the flare ribbons.\\
- We do not find signs of impact polarization above our noise level of $\sim$1\% in the 8542~\AA\ line. The only visible linear polarization signals show the typical symmetric Zeeman shape with a polarization of up to 10\%. However, we cannot exclude weak impact polarization signals which may be buried in the Zeeman signatures or signals that may have been present earlier, closer to flare maximum.\\
- Both flare ribbons are invisible below a certain height in the upper photosphere. Each of them crosses the umbra of one of the sunspots and they are the footpoints of a coronal loop system.\\
- The temporal evolution shows the flare ribbons becoming weaker in intensity and some structures disappearing. This may be caused by either a decreasing number of particles impacting from the corona or by heat dissipation in the chromosphere. While the brightness and vertical extent of the ribbon is probably dominated by the flare energy, we also find the ribbon structure evolving spatially.\\
- Sudden brightenings can be found inside the ribbons. They are followed by asymmetric profiles and probably simultaneous up- and downflows. They may be caused by a variable electron precipitation from the corona.

So how do our observation fit into the bigger flare picture? The field lines in the corona may have shifted and triggered a reconnection event, creating accelerated particle beams from the corona to the chromospheric ribbons. These beams were not energetic enough to permeate to the photosphere (as in white light flares), but only reached a certain height in the upper photosphere, at least during this declining phase. As the particle beam weakened, the flare ribbons became darker, with occasional energy releases because of enhanced particle beams or reconnections. Chromospheric motions (chromospheric evaporation) redistributed the energy and the magnetic fields, which did not reach photospheric levels. The continuing up- and inflow of material in the OPP may have built up more energy, which was released in a series of small flares after our observations stopped. Only when the field reached a more potential configuration the flare activity subsided and a penumbra was formed around the smaller spot.

Simultaneous high-resolution spectropolarimetric data of the photosphere and the chromosphere are valuable to investigate flare mechanisms and observations with imaging spectropolarimeters should continue to capture flares in all stages of evolution.

\acknowledgments

I am very grateful to the observers of the DST for obtaining great data. The observations were carried out as a visiting student at Sunspot, NM with kind help of H.~Uitenbroek. I thank P.~Judge for the discussions and K.~Reardon and A.~Tritschler for their help with the data reduction. I want to acknowledge the helpful discussions that took place during the ISSI meeting "Filamentary Structure and Dynamics of Solar Magnetic Fields", especially the suggestions by R.~Schlichenmaier, O.~Steiner and V.~Yurchyshyn. IBIS is a project of INAF/OAA with additional contributions from Univ.~of Florence and Rome and NSO.

\bibliographystyle{apj}
\bibliography{journals,ibisflare}

\begin{thebibliography}{37}
\expandafter\ifx\csname natexlab\endcsname\relax\def\natexlab#1{#1}\fi

\bibitem[{{Asai} {et~al.}(2004){Asai}, {Yokoyama}, {Shimojo}, {Masuda},
  {Kurokawa}, \& {Shibata}}]{asaietal2004}
{Asai}, A., {Yokoyama}, T., {Shimojo}, M., {et~al.} 2004, \apj, 611, 557

\bibitem[{{Aschwanden}(2002)}]{aschwanden2002}
{Aschwanden}, M.~J. 2002, \ssr, 101, 1

\bibitem[{{Benz}(2008)}]{benz2008}
{Benz}, A.~O. 2008, Living Reviews in Solar Physics, 5, 1

\bibitem[{{Bianda} {et~al.}(2005){Bianda}, {Benz}, {Stenflo}, {K{\"u}veler}, \&
  {Ramelli}}]{biandaetal2005}
{Bianda}, M., {Benz}, A.~O., {Stenflo}, J.~O., {K{\"u}veler}, G., \& {Ramelli},
  R. 2005, \aap, 434, 1183

\bibitem[{{Brooks} {et~al.}(2003){Brooks}, {Kurokawa}, {Yoshimura}, {Kozu}, \&
  {Berger}}]{brooksetal2003}
{Brooks}, D.~H., {Kurokawa}, H., {Yoshimura}, K., {Kozu}, H., \& {Berger},
  T.~E. 2003, \aap, 411, 273

\bibitem[{{Cavallini}(2006)}]{cavallini2006}
{Cavallini}, F. 2006, \solphys, 236, 415

\bibitem[{{Chen} {et~al.}(1994){Chen}, {Wang}, {Zirin}, \& {Ai}}]{chenetal1994}
{Chen}, J., {Wang}, H., {Zirin}, H., \& {Ai}, G. 1994, \solphys, 154, 261

\bibitem[{{Cheung} {et~al.}(2008){Cheung}, {Sch{\"u}ssler}, {Tarbell}, \&
  {Title}}]{cheungetal2008}
{Cheung}, M.~C.~M., {Sch{\"u}ssler}, M., {Tarbell}, T.~D., \& {Title}, A.~M.
  2008, \apj, 687, 1373

\bibitem[{{de la Cruz Rodr{\'{\i}}guez} \&
  {Socas-Navarro}(2011)}]{jaimehector2011}
{de la Cruz Rodr{\'{\i}}guez}, J., \& {Socas-Navarro}, H. 2011, \aap, 527, L8

\bibitem[{{Fletcher}(2005)}]{fletcher2005}
{Fletcher}, L. 2005, \ssr, 121, 141

\bibitem[{{Fletcher} \& {Hudson}(2001)}]{fletcherhudson2001}
{Fletcher}, L., \& {Hudson}, H. 2001, \solphys, 204, 69

\bibitem[{{Freeland} \& {Handy}(1998)}]{freelandhandy1998}
{Freeland}, S.~L., \& {Handy}, B.~N. 1998, \solphys, 182, 497

\bibitem[{{Giovanelli}(1978)}]{giovanelli1978}
{Giovanelli}, R.~G. 1978, \solphys, 59, 293

\bibitem[{{Handy} {et~al.}(1999){Handy}, {Acton}, {Kankelborg}, {Wolfson},
  {Akin}, {Bruner}, {Caravalho}, {Catura}, {Chevalier}, {Duncan}, {Edwards},
  {Feinstein}, {Freeland}, {Friedlaender}, {Hoffmann}, {Hurlburt}, {Jurcevich},
  {Katz}, {Kelly}, {Lemen}, {Levay}, {Lindgren}, {Mathur}, {Meyer}, {Morrison},
  {Morrison}, {Nightingale}, {Pope}, {Rehse}, {Schrijver}, {Shine}, {Shing},
  {Strong}, {Tarbell}, {Title}, {Torgerson}, {Golub}, {Bookbinder}, {Caldwell},
  {Cheimets}, {Davis}, {Deluca}, {McMullen}, {Warren}, {Amato}, {Fisher},
  {Maldonado}, \& {Parkinson}}]{handyetal1999}
{Handy}, B.~N., {Acton}, L.~W., {Kankelborg}, C.~C., {et~al.} 1999, \solphys,
  187, 229

\bibitem[{{Henoux} {et~al.}(1990){Henoux}, {Chambe}, {Smith}, {Tamres},
  {Feautrier}, {Rovira}, \& {Sahal-Brechot}}]{henouxetal1990}
{Henoux}, J.~C., {Chambe}, G., {Smith}, D., {et~al.} 1990, \apjs, 73, 303

\bibitem[{{H{\'e}noux} \& {Karlick{\'y}}(2003)}]{henouxkarlicky2003}
{H{\'e}noux}, J.-C., \& {Karlick{\'y}}, M. 2003, \aap, 407, 1103

\bibitem[{{Hudson}(2011)}]{hudson2011}
{Hudson}, H.~S. 2011, \ssr, 158, 5

\bibitem[{{Ichimoto} \& {Kurokawa}(1984)}]{ichimotokurokawa1984}
{Ichimoto}, K., \& {Kurokawa}, H. 1984, \solphys, 93, 105

\bibitem[{{Kosovichev} \& {Zharkova}(2001)}]{kosovichevzharkova2001}
{Kosovichev}, A.~G., \& {Zharkova}, V.~V. 2001, \apjl, 550, L105

\bibitem[{{Kurokawa}(1987)}]{kurokawa1987}
{Kurokawa}, H. 1987, \solphys, 113, 259

\bibitem[{{Li} \& {Zhang}(2009)}]{lizhang2009}
{Li}, L., \& {Zhang}, J. 2009, \apjl, 706, L17

\bibitem[{{Lim} {et~al.}(2010){Lim}, {Chae}, {Jing}, {Wang}, \&
  {Wiegelmann}}]{limetal2010}
{Lim}, E.-K., {Chae}, J., {Jing}, J., {Wang}, H., \& {Wiegelmann}, T. 2010,
  \apj, 719, 403

\bibitem[{{Metcalf} {et~al.}(1992){Metcalf}, {Wuelser}, {Canfield}, \&
  {Hudson}}]{metcalfetal1992}
{Metcalf}, T.~R., {Wuelser}, J.-P., {Canfield}, R.~C., \& {Hudson}, H.~S. 1992,
  in NASA Conference Publication, Vol. 3137, NASA Conference Publication, ed.
  {C.~R.~Shrader, N.~Gehrels, \& B.~Dennis}, 536

\bibitem[{{Murray} {et~al.}(2011){Murray}, {Bloomfield}, \&
  {Gallagher}}]{murrayetal2011}
{Murray}, S.~A., {Bloomfield}, D.~S., \& {Gallagher}, P.~T. 2011, \solphys, 129

\bibitem[{{Petrie} \& {Sudol}(2010)}]{petriesudol2010}
{Petrie}, G.~J.~D., \& {Sudol}, J.~J. 2010, \apj, 724, 1218

\bibitem[{{Reardon} \& {Cavallini}(2008)}]{reardoncavallini2008}
{Reardon}, K.~P., \& {Cavallini}, F. 2008, \aap, 481, 897

\bibitem[{{Scharmer} {et~al.}(2008){Scharmer}, {Narayan}, {Hillberg}, {de la
  Cruz Rodriguez}, {L{\"o}fdahl}, {Kiselman}, {S{\"u}tterlin}, {van Noort}, \&
  {Lagg}}]{scharmeretal2008}
{Scharmer}, G.~B., {Narayan}, G., {Hillberg}, T., {et~al.} 2008, \apjl, 689,
  L69

\bibitem[{{Schlichenmaier} {et~al.}(2010{\natexlab{a}}){Schlichenmaier}, {Bello
  Gonz{\'a}lez}, \& {Rezaei}}]{schlichenmaieretal2010b}
{Schlichenmaier}, R., {Bello Gonz{\'a}lez}, N., \& {Rezaei}, R.
  2010{\natexlab{a}}, ArXiv e-prints: 1009.4457

\bibitem[{{Schlichenmaier} {et~al.}(2010{\natexlab{b}}){Schlichenmaier},
  {Rezaei}, {Bello Gonz{\'a}lez}, \& {Waldmann}}]{schlichenmaieretal2010}
{Schlichenmaier}, R., {Rezaei}, R., {Bello Gonz{\'a}lez}, N., \& {Waldmann},
  T.~A. 2010{\natexlab{b}}, \aap, 512, L1

\bibitem[{{Schrijver}(2009)}]{schrijver2009}
{Schrijver}, C.~J. 2009, Advances in Space Research, 43, 739

\bibitem[{{Sheeley} {et~al.}(2004){Sheeley}, {Warren}, \&
  {Wang}}]{sheeleyetal2004}
{Sheeley}, Jr., N.~R., {Warren}, H.~P., \& {Wang}, Y.-M. 2004, \apj, 616, 1224

\bibitem[{{Su} {et~al.}(2011){Su}, {Jing}, {Wang}, {Mao}, {Wang}, {Zhang},
  {Deng}, {Guo}, \& {Wang}}]{suetal2011}
{Su}, J.~T., {Jing}, J., {Wang}, H.~M., {et~al.} 2011, \apj, 733, 94

\bibitem[{{{\v S}vestka} {et~al.}(1962){{\v S}vestka}, {Kopeck{\'y}}, \&
  {Blaha}}]{svestkaetal1962}
{{\v S}vestka}, Z., {Kopeck{\'y}}, M., \& {Blaha}, M. 1962, Bull. Astron. Inst.
  Czech., 13, 37

\bibitem[{{Vogt} \& {H{\'e}noux}(1999)}]{vogthenoux1999}
{Vogt}, E., \& {H{\'e}noux}, J.-C. 1999, \aap, 349, 283

\bibitem[{{Wang} {et~al.}(1994){Wang}, {Ewell}, {Zirin}, \&
  {Ai}}]{wangetal1994}
{Wang}, H., {Ewell}, Jr., M.~W., {Zirin}, H., \& {Ai}, G. 1994, \apj, 424, 436

\bibitem[{{Wittmann}(1971)}]{wittmann1971}
{Wittmann}, A. 1971, \solphys, 20, 365

\bibitem[{{W{\"o}ger} \& {von der L{\"u}he}(2008)}]{woegervdl2008}
{W{\"o}ger}, F., \& {von der L{\"u}he}, O. 2008, in Society of Photo-Optical
  Instrumentation Engineers (SPIE) Conference Series, Vol. 7019, Society of
  Photo-Optical Instrumentation Engineers (SPIE) Conference Series

\end{thebibliography}

\end{document}